\documentclass{article}
\usepackage{spconf,amsmath,graphicx}

\usepackage{amsfonts}
\usepackage{multirow}
\usepackage{hyperref}
\usepackage{subfigure}
\usepackage{booktabs,setspace}
\usepackage{threeparttable}
\ninept
\usepackage{setspace}

\usepackage{amssymb,CJK, indentfirst,balance,appendix}

\title{THE XMUSPEECH SYSTEM FOR MULTI-CHANNEL MULTI-PARTY MEETING TRANSCRIPTION CHALLENGE}
\name{\begin{tabular}{c}Jie Wang$^1$, Yuji Liu$^{3,4}$, Binling Wang$^2$, Yiming Zhi$^2$, Song Li$^1$, \\ 
Shipeng Xia$^2$, Jiayang Zhang$^{3,4}$, Lin Li$^{1}$, Qingyang Hong$^{2}$, Feng Tong$^{3,4}$\end{tabular}}

\address{$^1$School of Electronic Science and Engineering, Xiamen University, China\\
  $^2$School of Informatics, Xiamen University, China\\
  $^3$College of Ocean and Earth Sciences, Xiamen University\\
  $^4$Key Laboratory of Underwater Acoustic Communication and \\
  {Marine Information Technology (Xiamen University), Minister of  Education}\\
  {\{lilin,qyhong\}@xmu.edu.cn}}

\begin{document}

%\ninept
%
\maketitle

\begin{abstract}
This paper describes the system developed by the XMUSPEECH team for the Multi-channel Multi-party Meeting Transcription Challenge (M2MeT). For the speaker diarization task, we propose a multi-channel speaker diarization system that obtains spatial information of speaker by Difference of Arrival (DOA) technology. Speaker-spatial embedding is generated by x-vector and s-vector derived from Filter-and-Sum Beamforming (FSB) which makes the embedding more robust. Specifically, we propose a novel multi-channel sequence-to-sequence neural network architecture named Discriminative Multi-stream Neural Network (DMSNet) which consists of Attention Filter-and-Sum block (AFSB) and Conformer encoder. We explore DMSNet to address overlapped speech problem on multi-channel audio. Compared with LSTM based OSD module, we achieve a decreases of 10.1\% in Detection Error Rate (DetER). By performing DMSNet based OSD module, the DER of cluster-based diarization system decrease significantly form 13.44\% to 7.63\%. Our best fusion system achieves 7.09\% and 9.80\% of the diarization error rate (DER) on evaluation set and test set.
\end{abstract}


\section{Introduction}
As the application of speech signal processing becoming more and more popular, the technologies, such as Automatic Speech Recognition (ASR), speaker diarization are facing many challenges in real-world scenarios. In particular, meeting scenario is one of the most challenging and valuable because of its complexity and diversity, including overlapped speech, unknown number of speakers, reverberation, etc. Multi-channel Multi-party Meeting Transcription Challenge (M2MeT) \cite{yu2021m2met} focuses on addressing the “who speaks what at when” problem in real-world multi-speaker meetings. This challenge consists of two tracks, namely speaker diarization and multi-speaker ASR. The audios of M2MeT corpus named AliMeeting are collected by microphone array. 

\begin{figure}[htb]
  \centering
  \centerline{\includegraphics[width=8.5cm]{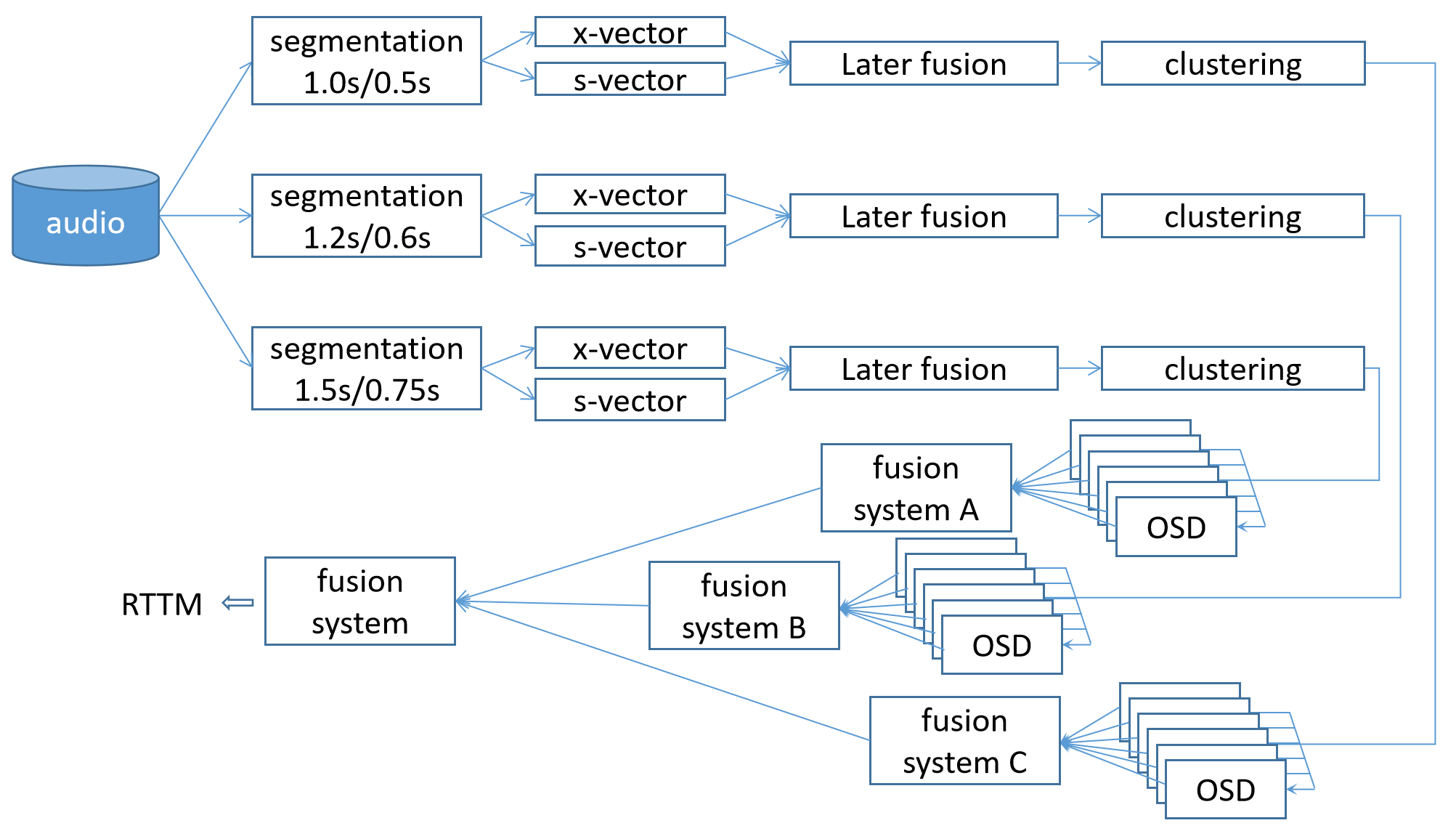}}
%  \vspace{2.0cm}
%  \centerline{}\medskip

\caption{Fusion speaker diarization system overview}
\label{fig:fusion_system}
\end{figure}

\begin{figure*}[!htb]
  \centering
  \centerline{\includegraphics[width=19cm]{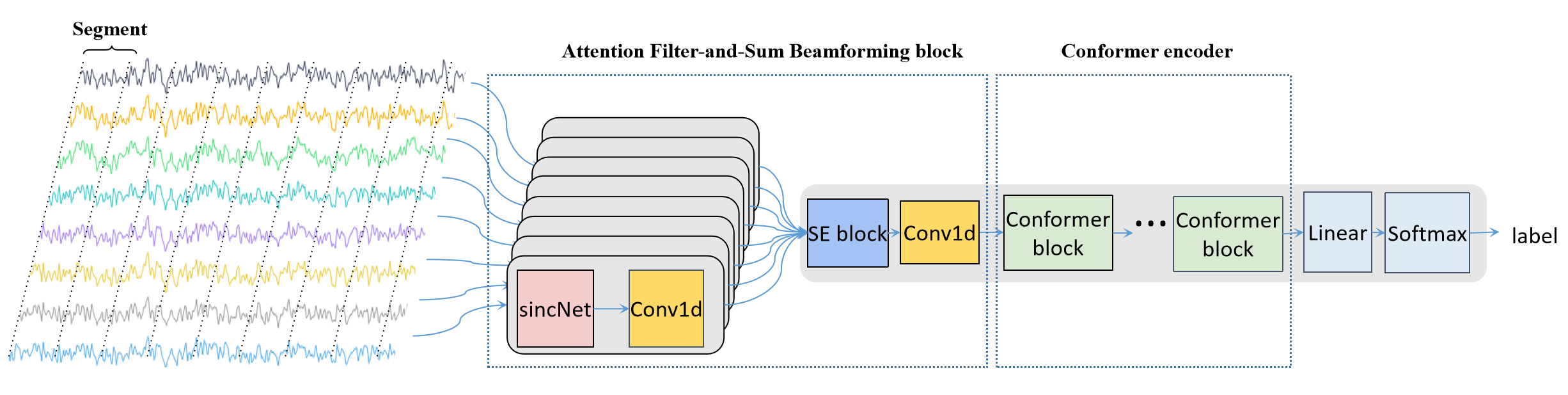}}
%  \vspace{2.0cm}
%  \centerline{}\medskip
\caption{DMSNet is a multi-channel sequence-to-sequence architecture used for sequence labeling.}
\label{fig:DMSNet}
\end{figure*}

One of the difficulties in M2Met is overlapped speech. There are a lot of overlapped regions in the meeting audios, which make this task more challenging. Overlapped speech not only cause wrong speakers number of clustering, but also increases the DER of system directly. Bredin H et al.in \cite{bredin2020pyannote} apply a Bi-LSTM based OSD named PyanNet to address single-channel overlapped speech. However, the audios of AliMeeting is 8-channel have rich spatial information, and the performance of Bi-LSTM based OSD still have room to improve. In this paper, inspired by filter-and-sum beamforming (FSB) algorithm\cite{frost1972algorithm}, we propose a novel architecture named Discriminative Multi-stream Neural Network (DMSNet) for overlapped speech detection. DMSNet is a novel multi-channel sequence-to-sequence architecture used for sequence labeling which consists of AFSB block and Conformer\cite{gulati2020conformer} encoder. Compared with Bi-LSTM based OSD model, DMSNet reduces Detection Error Rate (DetER) from 42.57\% to 32.47\%.What’s more, The audios of AliMeeting are collected from different rooms which is full of noise and reverberation. It is worth noting that the participants are required to remain in the same position during recording. In order to make full use of spatial information provided by the microphone array, we propose a multi-channel speaker diarization fusion system which combines spatial embedding and speaker embedding. In our systems, we perform Direction-of-Arrival (DOA) technology to extract spatial embedding, and combine x-vector to achieve better performance. The speaker diarization subsystem consists of speaker embedding extractor, spatial embedding extractor, clustering module, and OSD module. Compared with diarization system without OSD module, Applying DMSNet as OSD, the diarization error rate (DER) of speaker diarization system reduces from 13.44\% to 7.63\%. We fused different subsystems with different OSD modules and time scales to achieve better performance. Finally, we achieve 7.09\% and 9.80\% DER on AliMeeting evaluation set and test set.

\section{Data Preparation}
AliMeeting corpus supported by M2MeT contains 118.75 hours of speech data in total. The dataset includes 240 meeting audios collected by 8-channel microphone array, of which 212 audios are training set (Train), 8 audios are evaluation set (Eval) and the rest are test sets (Test). In particular, the average speech overlap ratio of training and evaluation set are 42.27\% and 34.76\%, respectively. We used the same systems which was trained with fixed training set in sub-track1 and sub-track2. AliMeeting, Aishell-4\cite{fu2021aishell} and CN-Celeb\cite{fan2020cn} as fixed training set can be used to train models. Our use of training set in this challenge is as follows:
\begin{itemize}
\item Speaker embedding extractor: We take CN-celeb1 (793 speakers) and CN-Celeb2 (2000 speaker) as the training set containing 2793 speakers in total (the CN-Celeb1-test part is excluded from training). 
\end{itemize}
\begin{itemize}
\item Overlapped speech detection: We use AliMeeting training set to train OSD models and evaluation set for validation.
\end{itemize}
\begin{itemize}
\item Data augmentation: We use the MUSAN\cite{snyder2015musan} (including music and noise) and RIRs \cite{ko2017study} to perform data augmentation.
\end{itemize}

\begin{table*}[]
\centering
\caption{The performance comparisons of OSD modules with different blocks on AliMeeting evaluation set. Detection Error Rate (DetER) is the main index which define the "miss error" and "false alarm error" DER.}
\label{tab:osd}
\begin{tabular}{llllllll}
\hline
ID & Architecture & Extract block & Encoder & DetER(\%) & Accuracy(\%) & Precision(\%) & Recall(\%) \\ \hline
M1 & PyanNet{\cite{bredin2020pyannote}} & SinConv(FSB) & Bi-LSTM & 42.57 & 91.61 & 85.23 & 70.06 \\
M2 & - & SinConv(FSB) & Conformer & 35.48 & 92.97 & 87.22 & 75.58 \\
M3 & - & AFSB(Shared) & Bi-LSTM & 36.50 & 92.72 & 87.47 & 74.12 \\
M4 & - & AFSB(Shared) & Conformer & 36.73 & 92.68 & 86.41 & 75.09 \\
M5 & - & AFSB(Discriminative) & Bi-LSTM & 36.68 & 92.69 & 88.18 & 73.12 \\
M6 & DMSNet & AFSB(Discriminative) & Confromer & \textbf{32.47 }& \textbf{93.53} & \textbf{89.22} &\textbf{ 76.81} \\ \hline
\end{tabular}
\end{table*}

\section{System Description}
\subsection{Task1 sub-track1\&2}
Our system speaker diarization system consists of speaker embedding extractor, spatial embedding extractor, clustering modules and OSD. The fusion system is illustrated in Figure \ref{fig:fusion_system}. We will describe each module in detail as follows:
\subsubsection{Segmentation }
M2MeT provide segments file to participants to get oracle VAD labels. The performance of fusion system which composed of subsystems with different time scale (including window length and time shift) will be improved. According to segments file we split segments of audio into different time scale.

\subsubsection{Speaker embedding}

The ResNet34-SE \cite{zhou2019deep} is employed as the x-vectors extractor with additive margin Softmax loss \cite{wang2018additive} which learns a segment-level representation from the input acoustic feature. The dimension of x-vector is reduced from 256 to 128 by Linear Discriminant Analysis (LDA). In our system, the input is 81-dimensional filter-banks extracted from the original 16kHz audio with a window size of 25ms and a 10ms shift.

\begin{figure}[htb]
  \centering
  \centerline{\includegraphics[width=8.5cm]{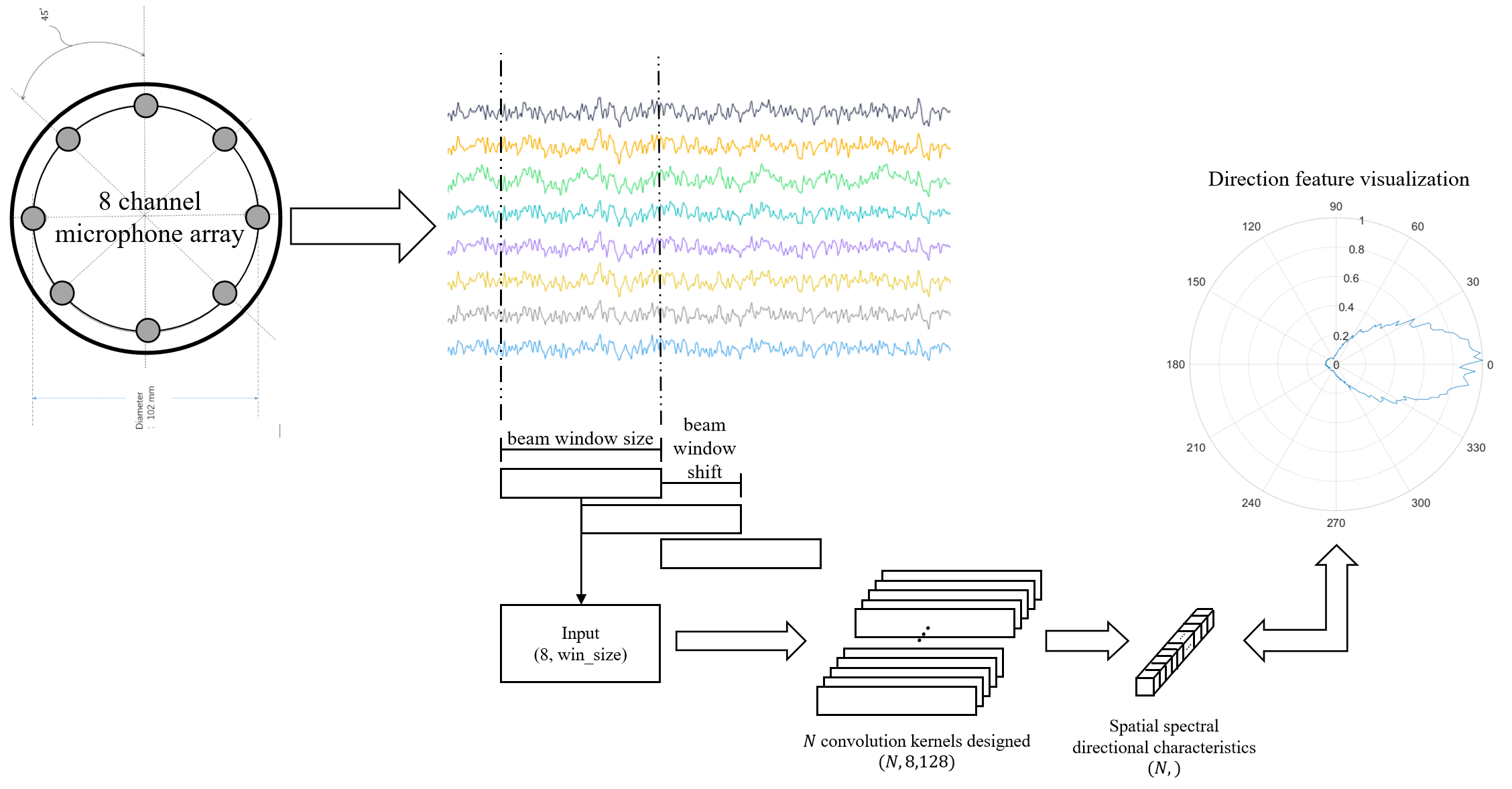}}
%  \vspace{2.0cm}
%  \centerline{}\medskip
\caption{A diagram of spatial embedding extractor.}
\label{fig:spatial_vector}
\end{figure}

\subsubsection{Spatial embedding}
The multi-channel signal provides spatial information compared to the single-channel signal. As show in Figure \ref{fig:spatial_vector}, FSB algorithm is used to extract spatial embedding named s-vectors. The signal of the \textit{M}-channel microphone array is denoted as \textit{x}$_{m}$(\textit{t})(\textit{m}=1,2,...,\textit{M}). We input the received signal through the corresponding Finite Impulse Response filter(FIR) and convert to single channel:
\begin{center}
\begin{equation}
y(t)=\sum_{m=1}^{M}\sum_{k=-K}^{K}a_{m,k}x_{m}(t-k)
\end{equation}
\end{center}
where \textit{K} is the order of the filter, and \textit{a}$_{m,k}$ is the \textit{k}$_{th}$ order weight of the \textit{m}$_{th}$ channel FIR filter \cite{zitouni2013simulated}.Convert Eq.1 to the frequency domain:
\begin{center}
\begin{equation}
\mathit{\mathbf{Y}}(e^{j\omega })=\mathit{\mathbf{F}}(\theta ,\omega )\mathit{\mathbf{S}}(e^{j\omega })
\end{equation}
\end{center}
where $\theta$ is the incident angle, and \textbf{F}($\theta$,$\omega$) is the spatial transmission response of the FSB algorithm in the direction $\theta$. The minimum mean square error (MMSE) criterion is used as the cost function to optimize the filter coefficients. Construct the following optimization problem
\begin{center}
\begin{equation}
\underset{a}{min}\iint\left| \mathbf{\mathit{F}}_{\mathbf{\mathit{d}}}(\theta-\varphi)-\mathbf{\mathit{F}}(\theta )\right|^2d\omega d\theta
\end{equation}
\end{center}
where $\varphi$ represents the angle away from the desired direction. \textbf{F}($\theta$,$\omega$) is denoted as:
\begin{center}
\begin{equation}
\mathit{\mathbf{F}}(\theta ,\omega )=\sum_{m=1}^{M}a_{m}z_{m}^{H}(\theta ,\omega )=\mathbf{a}\mathbf{z}^{H}(\theta ,\omega )
\end{equation}
\end{center}
Then the coefficients of spatial filter  can be obtained as:
\begin{center}
\begin{equation}
\mathbf{a}=\left (\sum_{l=1}^{L}\mathbf{z}_{l}\mathbf{z}_{l}^{H}  \right )^{-1}\left ( \mathbf{b}_{\varphi }\sum_{l=1}^{L}\mathbf{z}_{l}^{H} \right )
\end{equation}
\end{center}
where $\mathbf{z}_{l}=\left\{ z^{H}(\theta _{1},\omega _{l}), z^{H}(\theta _{2},\omega _{l}),\cdots ,z^{H}(\theta _{D},\omega _{l})\right\}$, $\mathbf{b}_{\varphi }=\left\{ \mathbf{F}_{d}(\theta _{1}-\varphi ),\mathbf{F}_{d}(\theta _{2}-\varphi ),\cdots ,\mathbf{F}_{d}(\theta _{N}-\varphi )\right\}$, \textit{L} and \textit{N} are the dimensions for dividing the frequency domain and the spatial domain, respectively.

In this paper, considering the complexity of the algorithm, 128-order filter coefficients are used, and 24, 36, 72, 120, 240, etc. are used for spatial domain division, respectively. The corresponding resolutions are 15°, 10°, 5°, 3° and 1.5°, the dimension of the filter is (\textit{N}, 8, 128), which is (\textit{N}, 8, 128) in this paper. This filter coefficient is placed in the neural network as an artificially designed \textit{N} convolution kernels of dimension (8, 128). As shown in Figure \ref{fig:spatial_vector}, when the 8-channel raw audio is sent to the network, single-channel signals in \textit{N} directions can be obtained. By normalizing the energy of signals in different \textit{N }directions, we can obtain spatial embedding named s-vector. The physical meaning is the spatial distribution of energy, and its statistical meaning is the probability of the existence of signals in \textit{N} directions. By increasing \textit{N}, higher resolution can be obtained in the physical sense, and more features can be seen from the perspective of neural network. Finally, more dimensional vectors are used for clustering, which can improve the robustness of s-vector.

\begin{table*}[]
\centering
\caption{DER(\%) of speaker diarization systems on evaluation set. M1$\sim$M6 are different OSD modules mention in section \ref{subsubsection:osd}}.
\label{tab:fusion}
\begin{tabular}{ccccccccccc}
\hline
ID & Time scales/s & Embedding & NME-SC & \multicolumn{6}{c}{OSD} & Fusion M1$\sim$M6 \\ \cline{5-10}
 &  &  &  & M1 & M2 & M3 & M4 & M5 & M6 &  \\ \hline
S1 & 1/0.5 & x-vector & 13.68 & 9.46 & 8.84 & 8.75 & 9.22 & 8.94 & \textbf{8.04} & 7.94 \\
S2 & 1.2/0.6 & x-vector & 13.53 & 9.16 & 8.41 & 8.25 & 8.69 & 8.40 & \textbf{7.84} & 7.73 \\
S3 & 1.5/0.75 & x-vector & 13.59 & 9.11 & 8.60 & 8.50 & 8.89 & 8.54 & \textbf{7.91} & 7.81 \\
S4 & 1/0.5 & sx-vector & 13.44 & 8.81 & 8.18 & 8.01 & 8.38 & 8.14 & \textbf{7.63} & 7.54 \\
S5 & 1.2/0.6 & sx-vector & 13.44 & 9.05 & 8.37 & 8.48 & 8.52 & 8.33 & \textbf{7.77 }& 7.63 \\
S6 & 1.5/0.75 & sx-vector & 13.57 & 9.09 & 8.50 & 8.45 & 8.92 & 8.51 & \textbf{7.81} & 7.64 \\ \hline
 & \multicolumn{9}{c}{Fusion S3 S4 S5} &\textbf{ 7.09} \\ \hline
\end{tabular}
\end{table*}

\subsubsection{Later fusion}
In the embedding fusion module, we perform late fusion method \cite{kang2020multimodal} to construct a separate similarity matrix (denoted below by \textbf{\textit{A$_{x}$}} and \textbf{\textit{A}}$_{s}$ for the x-vector and s-vector, respectively). We score the cosine similarity of embeddings in pairs to get similarity matrix. We use the following formula to yield the fused similarity matrix which is the input of clustering module. After tuning on evaluation set, we set the \textit{a} to 0.95.

\begin{center}
\begin{equation}
{\mathbf{A}}=a {\mathbf{A_{d}}} +(1-a) {\mathbf{A_{s}}}
\end{equation}
\end{center}

\subsubsection{Clustering}

In this stage, we perform normalized maximum eigengap spectral clustering (NME-SC) \cite{park2019auto} to obtain speaker labels of segments. NME-SC is a kind of spectral clustering algorithm, which can automatically estimate the number of clusters. During initial clustering, we found that if we tune the time scale of segments, the speaker diarization system will estimate the number of speakers more accurately. 

\subsubsection{Overlapped speech detection} \label{subsubsection:osd} 

The duration of overlapped speech accounts for a large proportion in AliMeeting corpus. In our system, we take two-stage OSD method to solve overlapped speech problem. The first stage is to detects overlapped region of speech. Then, In the second stage, we use the output of heuristic algorithm \cite{otterson2007efficient} to obtain secondary speaker labels. The secondary speaker labels of non-overlapped region will be removed. In the first stage, we propose a novel neural network architecture named DMSNet to detect overlapped speech of multi-channel audio. DMSNet is a sequence-to-sequence architecture which can be addressed as a sequence labeling task which matches a feature sequence X to the corresponding label sequence y where \textit{\textbf{X}} = {x$_1$,x$_2$,...,x$_T$} and label y = {y$_1$, y$_2$,...,y$_T$}. We use pyannote.audio toolkit \cite{bredin2020pyannote} to built this module. As shown in figure \ref{fig:DMSNet}, DMSNet consists of Attention Filter-and-sum Beamforming (AFSB) block and Conformer encoder. Inspired by FSB algorithm which design different artificial filters for channels to enhance the signal, we propose a learnable AFSB block to overcomes many shortcomings of the FSB. In AFSB block, the 8-channels raw audio is split into segments by sliding window and the feature of segments is extracted by SincNet \cite{ravanelli2018speaker} and convolutional layers. The weights of each channels are different, which make the module learn the spatial information of speech better. Squeeze-and-excitation (SE) block \cite{hu2018squeeze} is applied to learn the weights of each channel. Then, we use 1D convolution whose kernel size is 1×1 to reduce the number of channel to 1. We also adopt 24 layers Conformer as encoder to extract the speaker and spatial information of signal. Finally the classification layer including linear and softmax layers output the label. We used AliMeeting to train the module with Binary Cross Entropy (BCE).

As reported in Table \ref{tab:osd}, DMSNet achieve the best performance among these architecture. AFSB block is a necessary components of DMSNet. We perform FSB to enhance the 8-channel audio on AliMeeting evaluation set in module 1 and module 2. The weights of convolution layer in AFSB (shared) block is shared by each channel. Considering that the filter and time delay of each channel is discriminative in FSB, we make the weight of each channel different in AFSB (Discriminative), which improve the performance of OSD module by comparison between module 4 and module 6. We take Conformer (24 layers) as encoder to replace the function of Bi-LSTM and reduce the DetER of OSD module from 42.57\% to 35.48\%. After replacing PyanNet with DMSNet, the DetER of the model is reduced from 42.57\% to 32.47\%.

\subsubsection{Fusion method}
DOVER-Lap \cite{raj2021dover} is a overlap-aware fusion method for speaker diarization systems. We take two tricks to improve the performance of speaker diarization system. 
The first trick is called multi time scales subsystems fusion. If we fuse subsystems with different time scales, the DER of the final fusion system can be improved. The shorter the segments, the higher temporal resolution of diarization system. In contrast, if we increase the length of segments, the x-vector of segments will become more robust.The second trick is to fuse subsystems with different OSD module to improve the performance of fusion system. As show in Figure \ref{tab:fusion}, we performed those ticks in the final system, which make the final fusion diarization system achieve the lowest DER of all our systems.

\subsection{Task2 sub-track1\&2}
For the multi-speaker ASR track with restricted datasets, we use near-field and far-field data from the AliMeeting dataset, and AISHELL-4 as the training set with triple speed perturbation data augmentation. The multi-speaker speech recognition model uses the Serialized Output Training (SOT) algorithm with the same configuration as the baseline, and we achieve a CER of 29\% on the validation set.

\section{Results and Analysis}

The performance of our system on evaluation set is shown in Table \ref{tab:fusion}. A total of 6 different OSD modules were applied on speaker diarization systems. DMSNet based OSD module (M6) achieve the best performance of all OSD modules. By comparing S1 and S4, we found that the performance of sx-vector which combine spatial embedding and speaker embedding is better than x-vector on different time scales. We performance tow-step system fusion. In the first step, we fused sub-systems with different OSD modules and the results was improved. We further fused the S3, S4 and S5 whose time scales are different and the DER of system is reduced to 7.09\%. We performed the fused system on test set and achieved 9.80\% DER.

\section{Conclusions}
This paper described the XMUSPEECH system submitted to the M2MeT. We proposed a sequence-to-sequence architecture named DMSNet to detect overlapped speech of multi-channels audio. We designed a multi-channels speaker diarization system consists of ResNet34-SE based x-vector extractor, FSB based s-vector extractor, NME-SC module, and DMSNet based OSD. Finally, we fuse sub-systems with different OSD module and time scales to achieved 7.09\% and 9.80\% DER on AliMeeting evaluation set and test set.


\bibliographystyle{IEEEbib}
\bibliography{strings,refs}

\end{document}